\documentclass[english,article,twocolumn,showpacs,amsmath,amssymb,superscriptaddress,floatfix]{revtex4}
\usepackage[T1]{fontenc}
\usepackage[latin1]{inputenc}
\usepackage{graphicx}

\usepackage{appendix}

\makeatletter

\usepackage{bm}
\usepackage{pifont}
\usepackage{float}
\usepackage[capitalize]{cleveref}

\usepackage{babel}
\makeatother

\newcommand{\comment}[1]{}

\begin{document}

\preprint{}

\title{Microwave Spectroscopy of a Carbon Nanotube Charge Qubit}

\author{Z. V. Penfold-Fitch}
\affiliation{Cavendish Laboratory, University of Cambridge, Cambridge CB3 0HE, United Kingdom}

\author{F. Sfigakis}
\altaffiliation{Present address: The Institute for Quantum Computing, University of Waterloo, Canada.}
\affiliation{Cavendish Laboratory, University of Cambridge, Cambridge CB3 0HE, United Kingdom}

\author{M. R. Buitelaar}
\email{m.buitelaar@ucl.ac.uk}
\affiliation{London Centre for Nanotechnology, University College London, London WC1H 0AH, United Kingdom}
\affiliation{Department of Physics and Astronomy, University College London, London WC1E 6BT, United Kingdom}

\date{\today}

\begin{abstract}
Carbon nanotube quantum dots allow accurate control of electron charge, spin and valley degrees of freedom in a material which is atomically perfect and can be grown isotopically pure. These properties underlie the unique potential of carbon nanotubes for quantum information processing, but developing nanotube charge, spin, or spin-valley qubits requires efficient readout techniques as well as understanding and extending quantum coherence in these devices. Here, we report on microwave spectroscopy of a carbon nanotube charge qubit in which quantum information is encoded in the spatial position of an electron. We combine radio-frequency reflectometry measurements of the quantum capacitance of the device with microwave manipulation to drive transitions between the qubit states. This approach simplifies charge-state readout and allows us to operate the device at an optimal point where the qubit is first-order insensitive to charge noise. From these measurements, we are able to quantify the degree of charge noise experienced by the qubit and obtain an inhomogeneous charge coherence of 5 ns. We use a chopped microwave signal whose duty-cycle period is varied to measure the decay of the qubit states, yielding a charge relaxation time of 48 ns.
\end{abstract}

\pacs{73.63.Kv, 73.63.Fg, 73.23.Hk, 73.21.La, 03.67.Lx}

\maketitle

%
\section{Introduction}

Carbon nanotube quantum dots are of interest for quantum information processing because of the ability to accurately control electron charge, spin, and valley degrees of freedom in a material which is atomically perfect and has a well-understood electronic spectrum \cite{Laird1,Kuemmeth,Bulaev,Churchill1,Churchill2,Laird2,Viennot,Viennot2}. The relative ratio of $^{12}$C and $^{13}$C isotopes can be controlled during growth, allowing detailed studies of decoherence due to hyperfine coupling in $^{13}$C enriched nanotubes \cite{Churchill1,Churchill2,Fischer} or studies of $^{12}$C purified nanotubes for which long spin coherence times are expected \cite{Fischer}. The presence of spin-orbit interaction - a result of the nanotube curvature - allows electrical control of the electron spins \cite{Kuemmeth,Bulaev}, which can be read out in nanotube double quantum dots using spin-to-charge conversion \cite{Churchill1,Churchill2,Chorley2}. The valley degree of freedom offers further functionality, and coherent control of a nanotube valley-spin qubit was recently demonstrated \cite{Laird2}. The reported inhomogeneous coherence (or dephasing) times of the valley-spin qubits is limited to approximately 8 ns, however, which is tentatively attributed to hyperfine interaction and the susceptibility to charge noise of the devices. More generally, fast electrical manipulation and coupling of spin or spin-valley qubits relies on a mixing of charge and spin degrees of freedom - via a spin-orbit or exchange interaction - and charge noise ultimately sets their fidelity \cite{Laird2,Viennot}. A direct measurement of charge coherence in nanotube quantum dots and developing techniques to extend charge coherence is thus important for the application of carbon nanotube charge, spin, or valley-spin qubits in the field of quantum information processing.

In this work, we combine radio-frequency (rf) readout techniques with microwave spectroscopy of a carbon nanotube charge qubit. By driving transitions between the two charge qubit states, we are able to measure both the inhomogeneous charge coherence time $T^*_2$ and the charge relaxation time $T_1$. We find that $T^*_2 \sim 5$ ns and $T_1 \sim 48$ ns, limited by charge noise and dot-lead coupling, respectively.

\begin{figure*}
\includegraphics[width=175mm]{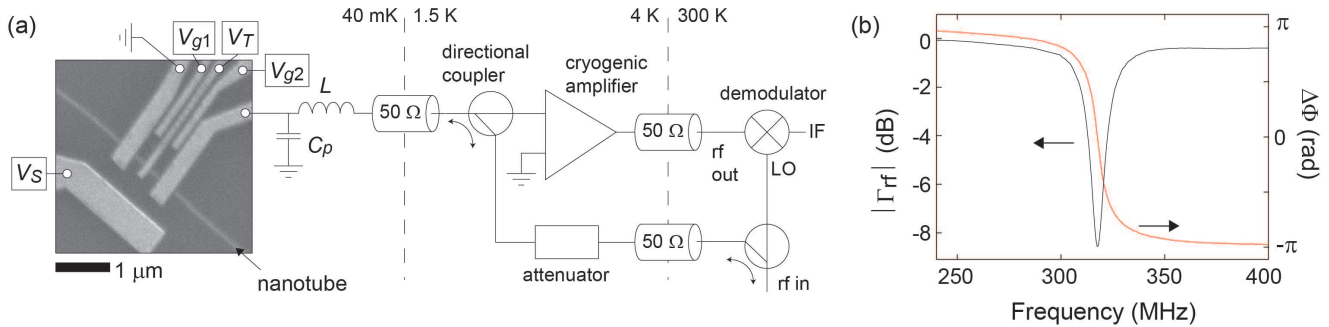}
\caption{\label{Fig1} \textbf{(a)} Schematic of the carbon nanotube device and the radio-frequency detection circuit. (Left panel) Scanning electron micrograph of a typical carbon nanotube double quantum dot device. The carbon nanotube is contacted by Au source and drain electrodes and capacitively coupled to an Al$_2$O$_3$/Ti/Au gate electrode that controls the tunnel barrier ($V_T$) between the quantum dots and three plunger gate electrodes that provide control of the left ($V_{g1}$) and right ($V_{g2}$) quantum dots, respectively, and overall control ($V_S$). (Right panel) The nanotube device is embedded in an \textit{LC} resonant circuit which both simplifies device fabrication and allows for high sensitivity and high bandwidth measurements. Room-temperature demodulation (LO, local oscillator and IF, intermediate frequency) provides a measurement of both quadratures. \textbf{(b)} Measured amplitude and phase response of a resonant circuit (background subtracted) with a resonant frequency of about 315 MHz. The precise resonant frequency depends on the quantum capacitance of the nanotube quantum dots, which can, therefore, be read out using this technique.}
\end{figure*}

\section{Radio-frequency reflectometry}

An important consideration for measurements of carbon nanotube devices is their one-dimensional geometry. While this geometry offers important advantages such as a natural confinement and large 'particle-in-a-box' energy scales, it also complicates the use of standard charge-state readout techniques such as proximal detectors. To overcome this barrier, we use a different method in which we measure the state-dependent quantum capacitance of a nanotube double quantum dot \cite{Chorley1} by coupling the device to a resonant electrical circuit as shown in Fig.~\ref{Fig1}. Key to the readout scheme is that the phase of a reflected rf signal depends on the quantum capacitance of the device and phase measurements thus provide a sensitive and noninvasive probe of the system.

This technique is illustrated in Fig.~\ref{Fig2}a, where we consider a double quantum dot near an effective $(1,0)-(0,1)$ charge transition, where the ordered pairs $(n,m)$ indicate the charge occupancies of the two quantum dots. In the presence of a tunnel coupling $t_c$ between the dots, these states form an effective two-level system that is described by the Hamiltonian $H=\frac{1}{2} \epsilon \hspace{0.5mm} \sigma_z + t_c \hspace{0.5mm} \sigma_x$, where $\epsilon$ is the energy difference, or detuning, between the two charge states in the absence of tunnel coupling. The quantum capacitance of the two-level system is directly proportional to the curvature - or second derivative - of the energy dispersion and has the largest magnitude at zero detuning where electrons move most easily between the two quantum dots. Importantly, the quantum capacitance has equal magnitude but opposite polarity for the bonding and antibonding states. These two states are thus distinguishable from phase measurements on the quantum dots using dispersive readout \cite{Chorley1,Petersson1,Colless1}.

\section{Carbon nanotube charge qubit}

The device we consider is a single-walled carbon nanotube on an undoped Si/SiO$_2$ substrate contacted by Au contacts. A central gate electrode is used to introduce a tunable tunnel barrier, separating the nanotube into two quantum dots, which can be individually addressed by two additional side gates; see Fig.~\ref{Fig1}a. A charge stability diagram is obtained from the rf phase response of the device as a function of the two side gates measured in a dilution refrigerator with electron temperature of about 80 mK as shown in Fig.~\ref{Fig2}b. In these measurements and those presented below, the rf frequency is set to the resonance frequency of the resonator where the phase sensitivity is largest; see Fig.~\ref{Fig1}b. The rf power at the device after attenuation is $P_{rf} \sim - 130$ dBm. From an analysis of the stability diagram (see Appendix \ref{appendix:b}) we are able to extract individual charging energies of about 6 meV for both dots and an electrostatic coupling energy between the dots of approximately 1.8 meV. The single-particle level spacing for nanotube quantum dots with a length of 500 nm is of order 2-3 meV. To lift the spin and valley degeneracies, we apply a magnetic field of 4 T with both perpendicular and parallel components to the nanotube axis. Here, we focus on an effective ($0,1$)-($1,0$) transition where the double quantum dot can be described as a quantum two-level system - the charge qubit \cite{Petta, Petersson2, Shi, Kim}.

\begin{figure*}
\includegraphics[width=170mm]{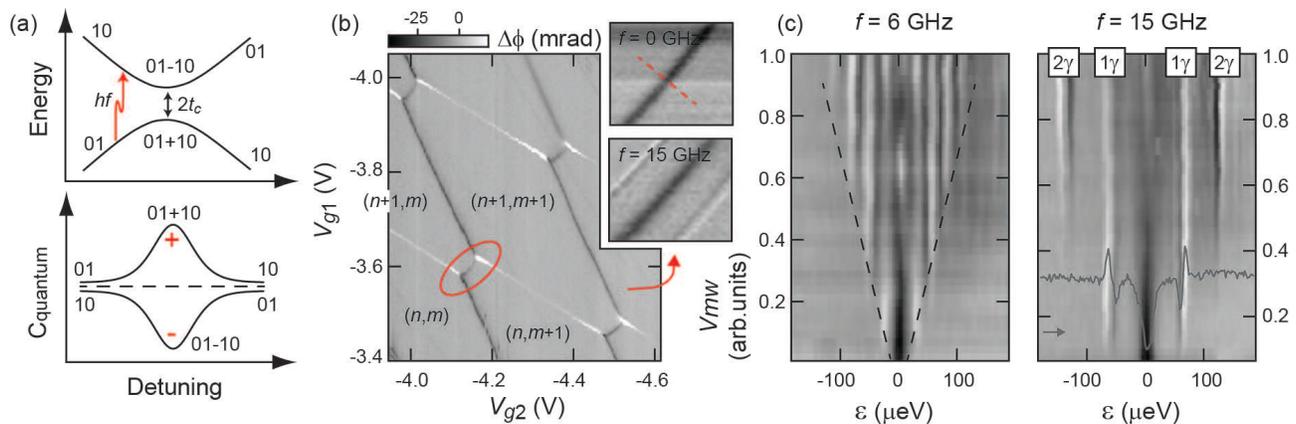}
\caption{\label{Fig2} \textbf{(a)} Energy band diagram (top panel) and corresponding quantum capacitance (bottom panel) as a function of level detuning for a double quantum dot with one electron. The indices denote the left and right dot occupancies \textbf{(b)} Measured phase response of a carbon nanotube double quantum dot coupled to an electrical resonator showing the charge stability diagram. The ordered pairs ($n,m$) indicate the effective charge occupancy. (Inset) Measured response along an effective (0,1)-(1,0) transition line in (top panel) the absence and (bottom panel) the presence of a 15-GHz microwave signal. \textbf{(c)} Phase response as a function of microwave amplitude across the charge transition - as indicated by the dashed red line in the inset of (b) - for a microwave frequency of (left panel) 6 and (right panel) 15 GHz. Multiphoton resonances ($\gamma = 0, 1, 2$) are observed for large amplitudes. The line trace shows the phase response at low microwave power as indicated by the arrow. The depth of the central dip is approximately 5 mrad.}
\end{figure*}

\subsection{Charge coherence time}

The phase signal measured at the charge transition line in the stability diagram [see inset to Fig.~\ref{Fig2}b] has a width and a height which are directly related to the tunnel coupling $t_c$ between the quantum dots \cite{Chorley1,Cottet}. To probe charge coherence, we apply - in addition to the rf readout - a microwave signal with frequency $f$ to one of the plunger gates such that the qubit is periodically driven across the anticrossing in the energy diagram. If the driving is sufficiently fast, there is the probability of a nonadiabatic transition from the ground to the excited state which can be understood as a Landau-Zener transition \cite{Shevchenko, Stehlik, Ribeiro, Cao}. As the system evolves, the ground and excited states acquire a phase difference and constructively interfere only if the energy difference between the states - which depends on the detuning - equals $nhf$, where $n$ is an integer and $h$ is Planck's constant. In other words, the qubit is resonantly driven between the ground and excited states only when their energy difference equals an integer number of the microwave-photon energy $hf$. In the stability diagram, this process is observed as lines parallel to the main transition line, as shown in the inset of Fig.~\ref{Fig2}b for $f=15$ GHz. We note that sidebands are observed for both positive and negative detuning, and asymmetric features associated with ($0,2$)-($1,1$) charge transitions \cite{Schreiber, Colless2} are not observed here.

The response of the device to the microwave drive is further illustrated in Fig.~\ref{Fig2}c where we show the interference pattern that is obtained when varying both the detuning [the dashed red line in the inset of Fig.~\ref{Fig2}b] and the applied microwave amplitude for frequencies of 6 and 15 GHz in the left and right panels, respectively. For large driving amplitudes, multiple-photon absorption is visible, with the lines most clearly separated in the 15-GHz data. These measurements also allow us to calibrate the microwave voltage at the device which can be obtained from the outline of the pattern (the dashed black lines). Importantly, the width of the resonances observed in Fig.~\ref{Fig2}c provide a measure of the inhomogeneous charge coherence time at finite detuning. We obtain a minimum width of approximately 8 $\mu$eV of the microwave-photon sidebands along the detuning axis, which corresponds to a frequency $\Delta f \sim 2$ GHz [as can also be seen directly in the measurements of Fig.~\ref{Fig3}a, right panel]. This yields an inhomogeneous charge coherence time $T^*_2 = 2 \sqrt{\ln 2}/ \pi \Delta f \sim 300$ ps.

\subsection{Measurements at optimal point}

\begin{figure*}
\includegraphics[width=165mm]{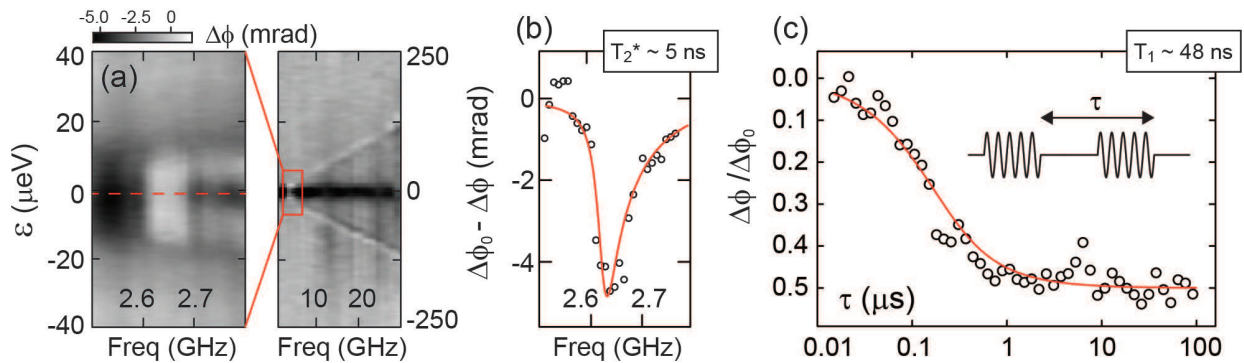}
\caption{\label{Fig3}\textbf{(a)} Phase response as a function of microwave frequency across the effective (0,1)-(1,0) charge transition. The sidebands are clearly visible at large applied frequencies. \textbf{(b)} At a frequency of approximately 2.65 GHz, the system is resonantly driven between the ground and excited states at zero detuning and a pronounced dip is observed in the phase response plotted as $\Delta \Phi_0 - \Delta \Phi$, where $\Delta \Phi_0 = -4.9$ mrad is the signal at zero detuning in the absence of microwave excitation. \textbf{(c)} Measurement of the charge relaxation time $T_1$ using chopped microwaves. The microwave signal is set to 2.65 GHz so that the qubit is driven resonantly between its ground and excited states. A sweep is taken along the line of detuning, and the length of the microwave pulse $\tau$ is stepped. A sweep across the zero-detuning line can be fitted to Eq.~\eqref{eq:2} to extract $T_1 = 48 \pm 6$ ns. (Inset) chopped microwaves with a 50$\%$ duty cycle.}
\end{figure*}

The most likely source of the short decoherence times observed at finite detuning is low-frequency charge noise in the detuning parameter. An advantage of the measurement technique used here is that we are able to directly measure at zero detuning where the qubit is first-order insensitive to charge noise \cite{Vion, Ithier}. These measurements are presented in Fig.~\ref{Fig3} and, indeed, show an increase of $T^*_2$ by more than 1 order of magnitude. The response of the phase signal in the presence of a microwave excitation for a wide frequency range up to 25 GHz is shown in the right panel of Fig.~\ref{Fig3}a. Visible is the phase signal at zero detuning as well as the $\gamma = 1$ photon sidebands. The separation of the sidebands increases with frequency and can be fitted accurately by the relation

\begin{equation}\label{eq:1}
\alpha V_g = \sqrt{(hf)^2 - (2t_c)^2}
\end{equation}

\noindent yielding an estimate of the tunnel coupling between the quantum dots of $t_c \approx 1.3$ GHz, and a conversion between gate voltage and detuning, $\epsilon = \alpha V_g$ with $\alpha \approx 0.01 |e|$.

A finely resolved measurement is shown in the left panel of Fig.~\ref{Fig3}a, and in a corresponding line trace at zero detuning in Fig.~\ref{Fig3}b for an applied microwave power $P_{mw} \sim -85$ dBm. A pronounced dip is observed at zero detuning at a frequency $f \sim 2.65$ GHz. The asymmetry of the dip is consistent with second-order charge noise at an optimal point: charge noise in the detuning parameter can only increase the energy splitting between the two eigenstates at zero detuning, resulting in a broader tail at higher frequencies \cite{Ithier}. We empirically fit the data assuming a Gaussian noise distribution with standard deviation $\sigma_{\epsilon} \approx 2.5$ $\mu$eV in the detuning parameter - consistent with estimates of $\sigma_{\epsilon}$ at finite detuning - which we convolve with a Lorentzian, the width of which is taken as a fit parameter, to account for charge relaxation (see Appendix \ref{appendix:c1} for further details).

A best fit to the data yields a Lorentzian linewidth of $\sim 30$ MHz which is consistent with the measured $T_1$ (see below). From the overall line shape [see the red solid line in Fig.~\ref{Fig3}b], we obtain $T^*_2 \sim 5$ ns. The observed inhomogeneous charge coherence time is an order of magnitude longer than observed in other nanotube experiments \cite{Viennot2} - a result of the suppression of first-order charge noise and a relatively long charge relaxation time which would otherwise obscure the asymmetry seen in Fig.~\ref{Fig3}b. We note that increasing the tunnel coupling $t_c$ between the quantum dots makes the qubit less susceptible to charge noise, but at the expense of a decrease in readout sensitivity \cite{Cottet, Chorley1}.

\subsection{Charge relaxation time}

To directly measure the charge relaxation time $T_1$ of the carbon nanotube charge qubit, we use chopped microwaves with a 50$\%$ duty cycle \cite{Petta}. The chopped microwave signal is created by combining a continuous microwave signal set to 2.65 GHz with a variable-width pulse provided by an arbitrary waveform generator. By varying the period $\tau$ - starting at $\tau = 15$ ns - we are able to measure charge relaxation as shown in Fig.~\ref{Fig3}c. During the first half of a cycle, the microwave signal is present and the qubit is resonantly driven between the two qubit states - followed by decay to the ground state during the second half when the microwave signal is switched off.

For very short timescales where $\tau \ll T_1$, the system has no time to relax and the averaged phase signal receives equal contributions from both qubit states (saturation). As these states have opposite polarity, the phase signal is close to the background value, which is set to zero here (see Appendix \ref{appendix:c2} for further details). For large time scales where $\tau \gg T_1$, the system has time to relax when the microwave signal is switched off. In this limit, the phase signal is expected to approach $\Delta \Phi = \Delta \Phi_{0}/2$ where $\Delta \Phi_{0}$ is the amplitude of the phase signal at zero detuning in the absence of any microwave excitation, i.e. the system at thermal equilibrium.

We fit the data measured at zero detuning by an exponential decay with a single fit parameter $T_1$:

\begin{equation}\label{eq:2}
\frac{\Delta \Phi (\tau)}{\Delta \Phi_{0}} = \frac{1}{2}-\frac{T_1(1-e^{-\tau/2T_1})}{\tau}
\end{equation}

\noindent and obtain a good fit with theory yielding $T_1 = 48 \pm 6$ ns. The charge relaxation mechanism can be both intrinsic - e.g. via electron-phonon coupling - or extrinsic to the carbon nanotube, such as relaxation via the leads \cite{Vorontsov}. We believe that, in our device, the mechanism is extrinsic and $T_1$ is limited by dot-lead coupling. A rough estimate of the coupling of the quantum dot to the  leads is obtained from standard dc measurements: when a dc bias voltage is applied over the double quantum dot, we are able to measure a small cotunneling current in the picoampere range which corresponds to a rate consistent with the observed charge relaxation time (see Appendix \ref{appendix:c2}). The observed $T_1$ should, therefore, be considered as a lower bound for intrinsic charge relaxation times in carbon nanotube quantum dots.

\section{Discussion and conclusion}

The results presented here show measurements of charge coherence and relaxation in a carbon nanotube charge qubit using a noninvasive readout technique that does not require proximal detectors. We are able to significantly extend charge coherence by operating the device at an optimal point, where it is first-order insensitive to charge noise. We observe $T^*_2\sim 5$ ns, which provides a benchmark for charge coherence in carbon nanotubes and which is comparable to, or longer than, charge coherence observed in GaAs- and Si-based devices\cite{Petta, Petersson2,Shi,Kim}. To extend coherence times further requires the improvement of device fabrication.  A specific advantage of carbon nanotubes in this respect is that they can be grown or placed on a wide range of substrate materials. Since impurities in - or polar adsorbates on - the SiO$_2$ or Al$_2$O$_3$ dielectrics are a likely source of charge noise, it would be of considerable interest to extend our measurements to substrates such as hexagonal boron nitride \cite{Baumgartner} or to ultraclean suspended carbon nanotube quantum dots \cite{Waissman, Ranjan}. The latter system would also be of interest as a model system for investigating coupling of charge and vibrational degrees of freedom \cite{Benyamini} and qubit coupling via the exchange of virtual phonons \cite{Wang}.

\section*{ACKNOWLEDGMENTS}

We thank Markus Weiss, Andreas Baumgartner and Christian Sch\"{o}nenberger for the carbon nanotube growth, Simon Chorley, James Frake and Jon Griffiths for the experimental assistance, and Joachim Wabnig for the useful discussions. We gratefully acknowledge funding from the European Research Council (ERC), Consolidator Grant Agreement No. 648229 (CNT-QUBIT).

\renewcommand{\appendixname}{APPENDIX}
\appendix

\section{Device fabrication and experimental methods}
\label{appendix:a}

Single-walled carbon nanotubes were grown by chemical vapor deposition on nominally undoped SiO$_2$ substrates with 280-nm thermal oxide. The room-temperature resistivity of the Si wafers is $\rho >100$ ohm cm. The carbon nanotubes have a natural abundance carbon isotope ratio (98.9\% $^{12}$C), and they are distributed at a concentration of approximately one nanotube per 10 ${\mu m}^2$ on the substrate. Device fabrication consists of three proximity-corrected electron-beam lithography and metal-evaporation stages using 5\% (by weight) of 495000-molecular-weight polymethyl methacrylate dissolved in anisole as a resist. During the first stage, alignment marks and bond pads are fabricated on the substrates using 5/55 nm of Ti/Au. The carbon nanotubes are subsequently located with respect to the alignment marks using atomic force microscopy and scanning electron microscopy. During the second lithography stage, the source and drain electrodes as well as the two plunger gates are defined using 50-nm Au. During the third lithography stage the central gate electrode is defined. Metal evaporation consists of the deposition of 1- to 2-nm Al followed by exposure to air for five min to oxidize the Al. A further evaporation-air-exposure cycle is carried out and, finally, 5/40 nm of Ti/Au is evaporated onto the sample. This process results in a thin insulating layer of Al$_2$O$_3$ between the carbon nanotube and the top gate. A wedge bonder is used to connect the large bond pads of the sample to the sample holder dc and rf gates using Au wires.\\

\medskip

Experiments are carried out in a dilution refrigerator with an electron temperature of about 80 mK. The measurement circuit includes several dc lines which are thermally anchored and extensively filtered at various stages. The radio-frequency detection circuit is connected to the device as shown in the schematics of Fig.~\ref{Fig1}a. Briefly, an attenuated radio-frequency signal is directed to the device source electrode through the coupled port of a directional coupler. Using the directional coupler, the reflected signal is extracted. The signal is first amplified by a cryogenic preamplifier anchored at 4 K, followed by room-temperature amplification. Demodulation is achieved by mixing the reflected signal with the reference signal. Both quadratures of the signal are detected, allowing measurements of both the amplitude and the phase response. The value of the chip inductor $L$ mounted on the sample holder is chosen for a resonance frequency $f_0 = 1/(2 \pi \sqrt{LC})$ in the (300 - 500)-MHz range, the operating range of the cryogenic amplifier. The noise temperature of the amplifier is $T_N \sim 2.8$ K at 350 MHz. The capacitance $C \sim 0.5$ pF is dominated by the parasitics of the sample holder and the device substrate which could be significantly reduced using undoped Si/SiO$_2$ as compared to doped Si/SiO$_2$ substrates, at the expense of back-gate tuning. For all measurements presented, the rf signal is turned on and set to be on resonance ($f_0$) where the response is most sensitive to changes in phase.\\

\medskip

\begin{figure}[b]
\includegraphics[width=85mm]{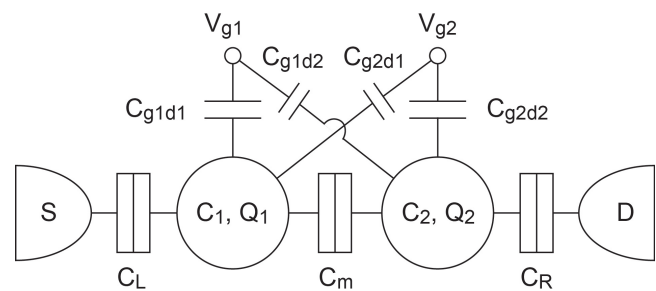}
\caption{\label{Fig4} Schematic capacitor model of the double quantum dot system with the various capacitances between quantum dots and electrodes indicated. The quantum dots are coupled to the source ($S$) and drain ($D$) electrodes via tunnel junctions.}
\end{figure}

\begin{figure*}
\includegraphics[width=140mm]{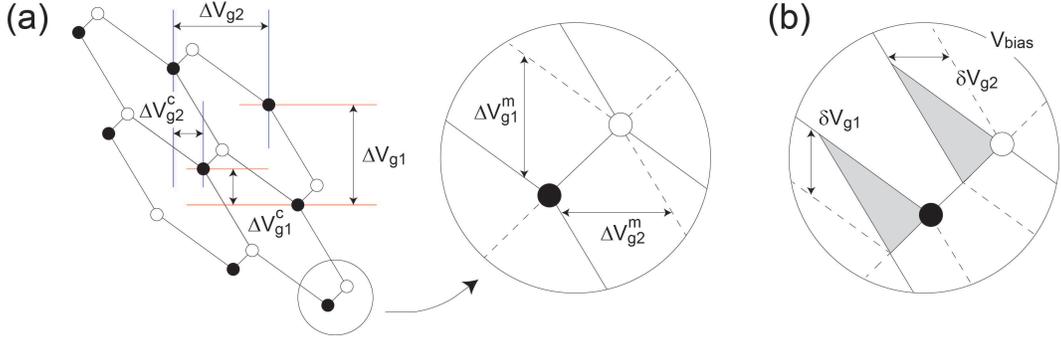}
\caption{\label{Fig5} (a) Schematic stability diagram showing the gate-voltage spacings used to determine the double dot capacitances in \cref{eq:b5,eq:b6,eq:b7}. (b) Bias triangles in the stability diagram in the presence of a source-drain voltage $V_{bias}$.}
\end{figure*}

A bias tee allows for both rf and dc signals to be applied to the device source electrode. The bias tee consists of a resistor, a capacitor and an inductor, with the values $R_B$ = 1 k$\Omega$, $C_B$ = 100 pF, and $L_B$ = 470 nH. The sample holder incorporates subminiature push-on connectors directed to the device plunger gates via microstrip lines for fast gating, enabling microwave spectroscopy. The microwave generator produces continuous-wave signals up to 40 GHz which are added to the dc gate signal via a broadband (Anritsu K251) bias tee. The chopped microwave signal is created by combining the continuous-wave signal with a variable-width pulse provided by an arbitrary-waveform generator using a microwave switch. For short pulse lengths, the experiment is limited by the approximately 5-ns risetime of the switch. All other electrodes used for dc gating are rf-grounded at the sample holder using 100-pF capacitances.

\section{Double Quantum Dot Stability Diagram}
\label{appendix:b}

\noindent We first consider a purely electrostatic model of the double quantum dot. The capacitor network of the model is shown in Fig.~\ref{Fig4}. In our analysis, we follow the work of van der Wiel \textit{et al} \cite{Wiel}, but we include cross-capacitances between gate 1(2) and quantum dot 2(1) which account for the skewing observed in the measured stability diagrams. The labeling of the various capacitances in the system is as indicated in the figure. The electrostatic energy of this system with $n$ and $m$ electrons on dots 1 and 2, respectively is then, up to an offset independent of $n$ and $m$, given by \cite{Wiel, Yamahata, Graber1}

\begin{equation}\label{eq:b1}
E_{n,m} = E_{C1}/2~n^2 + E_{C2}/2~m^2 + E_{Cm} nm + E_1 n + E_2 m
\end{equation}

where $E_{C1(2)}$ represents the charging energies of the two dots, $E_{Cm}$ is the electrostatic coupling energy and $E_1$ and $E_2$ are the single-particle energies of the dots provided by the two gate electrodes \cite{Graber1}. Defining the sum of the capacitances directly coupled to each quantum dot as $C_{1(2)} = C_{L(R)} + C_{g1(2)d1(2)} + C_{g2(1)d1(2)} + C_m$, the energies are given by:

\begin{widetext}
\begin{equation}\label{eq:b2}
E_{C1}=e^2 \frac{C_2}{C_1 C_2 - C_m^2}; \quad E_{C2}=e^2 \frac{C_1}{C_1 C_2 - C_m^2}; \quad E_{Cm}=e^2 \frac{C_m}{C_1 C_2 - C_m^2}
\end{equation}
\end{widetext}

Note that $E_{C1}$ and $E_{C2}$ correspond to charging energies $e^2/C_{\Sigma}$ for dots 1 and 2, where $C_{\Sigma}$ equals $C_{1(2)}$ of the individual dots with a correction factor due to the coupling; i.e. $C_{\Sigma 1(2)} = C_{1(2)} - C^2_m / C_{2(1)}$ The energies $E_1$ and $E_2$ are \cite{Wiel, Yamahata}

\begin{figure*}
\includegraphics[width=152mm]{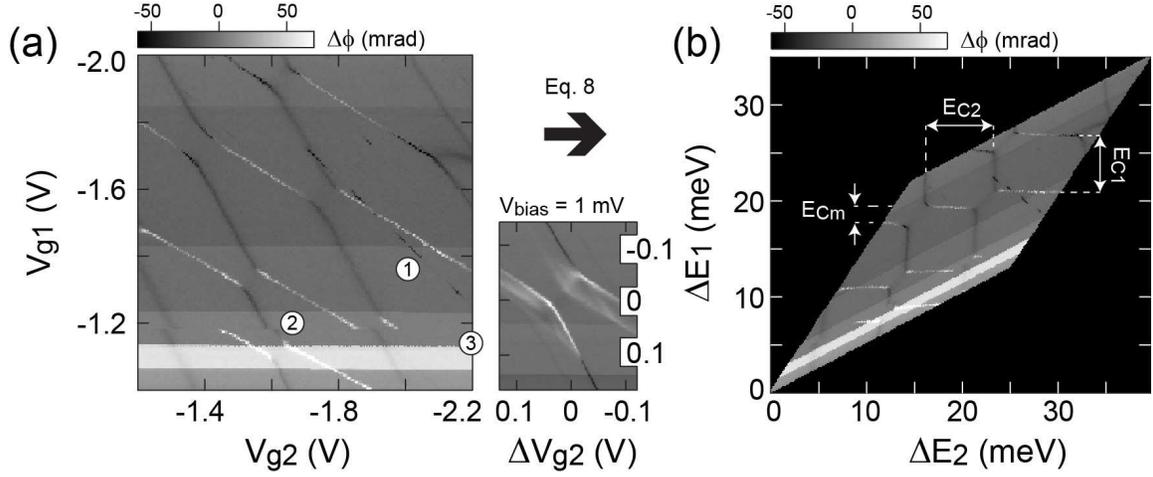}
\caption{\label{Fig6} (a) Measured stability diagram showing the phase response as a function of the gate voltages $V_{g1}$ and $V_{g2}$. The magnetic field $B=0$ T. The numbers 1-3 indicate different types of measurement artifacts described in the text. (Inset) The phase response of bias triangles for $V_{bias}=1$ mV. (b) Measured stability diagram of the data in (a) shown as a function of $E_1$ and $E_2$ using Eq.~\eqref{eq:b8} to correct for the skewing.}
\end{figure*}

\begin{widetext}
\begin{align}\label{eq:b3}
\begin{split}
E_1 & = A V_{g1} + B V_{g2} = -(C_{g1d1} E_{C1} + C_{g1d2} E_{Cm})V_{g1}/|e| - (C_{g2d2} E_{Cm} + C_{g2d1} E_{C1})V_{g2}/|e|\\
E_2 & = C V_{g1} + D V_{g2} = -(C_{g1d1} E_{Cm} + C_{g1d2} E_{C2})V_{g1}/|e| - (C_{g2d2} E_{C2} + C_{g2d1} E_{Cm})V_{g2}/|e|
\end{split}
\end{align}
\end{widetext}

\noindent where $A,B,C$,and $D$ are conversion factors between gate voltage and energy for use in Eq.~\eqref{eq:b8} below. The boundaries of the stability diagram then follow from the electrochemical potentials of the two quantum dots:
\begin{widetext}
\begin{align}\label{eq:b4}
\begin{split}
\mu_1(n,m) & = E_{n,m}-E_{n-1,m} = (n-1/2)E_{C1} + m E_{Cm} + E_1\\
\mu_2(n,m) & = E_{n,m}-E_{n,m-1} = (m-1/2)E_{C2} + n E_{Cm} + E_2
\end{split}
\end{align}
\end{widetext}

\noindent which allows us to calculate the various capacitances from measurements of the gate-voltage spacing between different triple-point pairs as indicated in Fig.~\ref{Fig5}a. The relations $\mu_{1,2}(n,m,V_{g1},V_{g2})=\mu_{1,2}(n+1,m,V_{g1}+\Delta V_{g1},V_{g2}-\Delta V^c_{g2})$ and $\mu_{1,2}(n,m,V_{g1},V_{g2})=\mu_{1,2}(n,m+1,V_{g1}-\Delta V^c_{g1},V_{g2}+\Delta V_{g2})$ using Eqs.~\eqref{eq:b3} and \eqref{eq:b4} then yield the capacitances between between gate 1(2) and dot 1(2) and the cross-capacitances between gate 1(2) and dot 2(1):

\begin{align}\label{eq:b5}
\begin{split}
C_{g1(2)d1(2)}& = \frac{\Delta V_{g2(1)}}{\Delta V_{g1}\Delta V_{g2}-\Delta V^c_{g1}\Delta V^c_{g2}}|e|\\
C_{g1(2)d2(1)}& = \frac{\Delta V^c_{g2(1)}}{\Delta V_{g1}\Delta V_{g2}-\Delta V^c_{g1}\Delta V^c_{g2}}|e|
\end{split}
\end{align}

\noindent Measurements of the gate-voltage spacing between triple points within a pair $\Delta V^m_{g1,2}$ allow us to calculate the ratio between $C_{1(2)}$ and $C_m$ using $\mu_1(n,m,V_{g1},V_{g2})=\mu_1(n,m+1,V_{g1}+\Delta V^m_{g1},V_{g2})$ and $\mu_2(n,m,V_{g1},V_{g2})=\mu_2(n+1,m,V_{g1},V_{g2}+\Delta V^m_{g2})$ as follows:
\begin{equation}\label{eq:b6}
\frac{C_{1(2)}}{C_m} = \frac{|e|}{C_{g2(1)d2(1)}\Delta V^m_{g2(1)}} - \frac{C_{g2(1)d1(2)}}{C_{g2(1)d2(1)}}
\end{equation}
\\
\noindent Finally, all capacitance values can be obtained by applying a finite source-drain bias voltage $V_{bias}$ across the double dot, see Fig.~\ref{Fig5}b, using the relations

\begin{align}\label{eq:b7}
\begin{split}
|e|\delta V_{g1}& =\frac{C_1 C_2 - C^2_m}{C_{g1d1} C_2 + C_{g1d2} C_m}|eV_{bias}|\\
|e|\delta V_{g2}& =\frac{C_1 C_2 - C^2_m}{C_{g2d2} C_1 + C_{g2d1} C_m}|eV_{bias}|
\end{split}
\end{align}

\subsection{Measured stability diagram}

\begin{figure*}
\includegraphics[width=125mm]{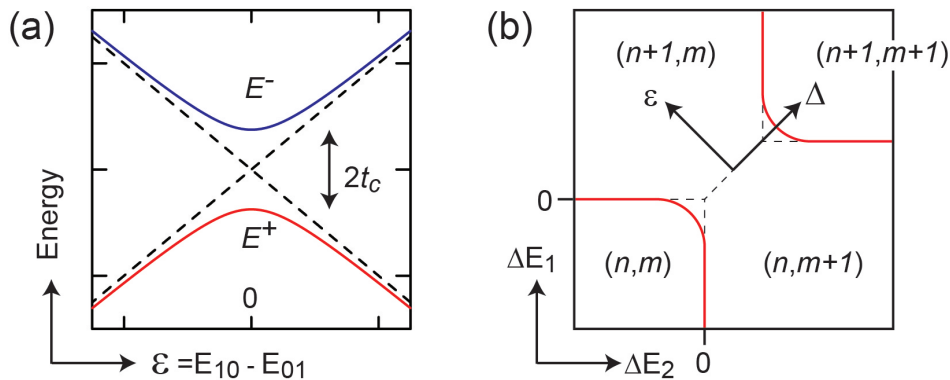}
\caption{\label{Fig7} (a) Energies of the bonding (red curve) and antibonding (blue curve) states as a function of detuning of the quantum two-level system described in the text. At zero detuning the two states are separated by an energy $2t_c$. (b) Schematic stability diagram. In the presence of a tunnel coupling the triple points develop into curved wings.}
\end{figure*}

\noindent A section of the stability diagram of the double dot device discussed in the main text is shown in Fig.~\ref{Fig6}a, in which we plot the phase response as a function of the two gate voltages $V_{g1}$ and $V_{g2}$. Using \cref{eq:b5,eq:b6,eq:b7} above, we are able to extract the various capacitances of the system: $C_{g1d1}=0.52$ aF,  $C_{g1d2}=0.21$ aF,  $C_{g2d2}=0.55$ aF,  $C_{g2d1}=0.18$ aF, $C_1=28.67$ aF,  $C_2=25.97$ aF and  $C_m=7.71$ aF. These capacitance values correspond to charging energies $E_{C1} = 6.06$ meV, $E_{C2} = 6.69$ meV and $E_{Cm} = 1.8$ meV. Using the matrix parameters $A-D$ defined in Eq.~\eqref{eq:b3} we are now able to provide a conversion from gate voltage to energy which yields
\begin{equation}\label{eq:b8}
\begin{pmatrix}
E_1 \\
E_2\\
\end{pmatrix}
 =
\begin{pmatrix}
0.0219 & 0.0130 \\
0.0147 & 0.0249 \\
\end{pmatrix}
\begin{pmatrix}
|e| V_{g1} \\
|e| V_{g2} \\
\end{pmatrix}
\end{equation}

\noindent and allows us to plot the experimental data as a function of $E_1$ and $E_2$; see Fig.~\ref{Fig6}b. We note that the experimental data show a variation in the size of the cells in the stability diagram reflecting the discrete energy spectrum of the quantum dots, which is not taken into account in the electrostatic model above. For nanotubes with a length $L \sim 500$ nm the single-particle energy spacing $h v_F/4L$ is on the order 2 meV if valley degeneracy is broken; using a Fermi velocity $v_F = 8\cdot 10^5$ m/s. The variation in the size of the cells (e.g. an even-odd periodicity reflecting spin degeneracy at zero field) nevertheless allows us to determine the charging energies and capacitances and provides an estimate of the single-particle energy spacing consistent with the length of the nanotube quantum dots.\\
\\
\noindent Figure ~\ref{Fig6}a also shows three different types of artifacts observed in the experiments:

(1): Additional lines in the stability diagram apparent in the measured rf response but not in the dc transport data. These resonances are superposed on the double quantum dot resonances and do not show signatures of interaction such as anticrossing. These additional lines could be due, e.g., to other nanotubes or oxide charge traps coupled to the rf electrode but not the nanotube quantum dot device.

(2): Sudden, but mostly reproducible, shifts in the stability diagram at specific gate voltages. These shifts are most likely due to charge traps in the SiO$_2$ or AlO$_x$ oxides not coupled to the rf electrode.

(3): Sudden shifts in the background phase signal at random times. These nonreproducible shifts - not observed for the data discussed in the main text - are unrelated to the double dot device but are due to the measurement circuitry and do not affect the analysis.\\

\begin{figure*}
\includegraphics[width=160mm]{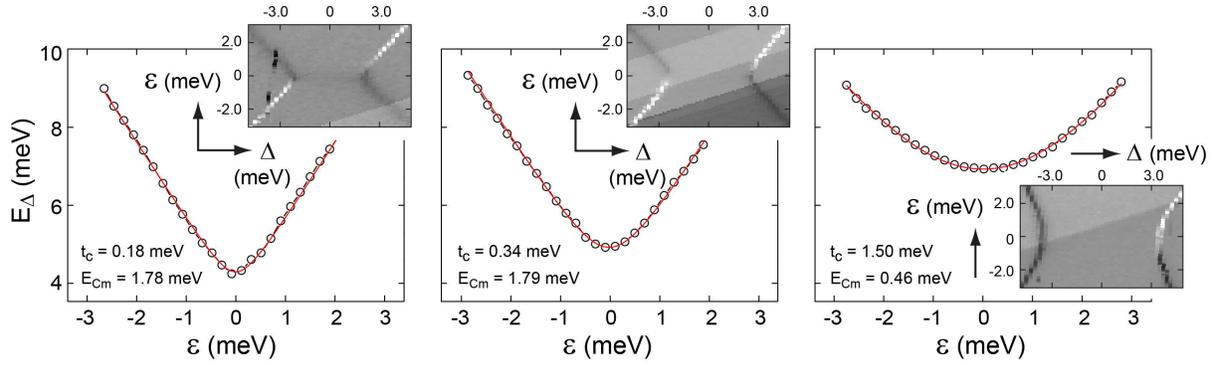}
\caption{\label{Fig8} Wing spacing $E_{\Delta}$ as a function of detuning $\epsilon$ for three different charge configurations with increasing tunnel couplings $t_c$ from left to right. The red lines show fits to the data using Eq.~\eqref{eq:b11}. (Inset) The corresponding stability diagram sections as a function of $\Delta = \Delta E_1 + \Delta E_2$ and $\epsilon = \Delta E_1 - \Delta E_2$ as described in the text.}
\end{figure*}

\subsection{Tunnel coupling}

The presence of a tunnel coupling $t_c$ between the two quantum dots results in the formation of bonding and antibonding states, which modifies the stability diagram: the triple points develop into curved lines, or wings, as evident in the data of Figs.~\ref{Fig6} and ~\ref{Fig8}. The basic features of the experiment are well described by a simple model of a quantum two-level system for a single electron in the double quantum dot, neglecting spin and interaction with electrons at lower energy levels \cite{Wiel}. Including the tunnel coupling $t_c$ between the quantum dots, the eigenstates of the quantum two-level system - our charge qubit - are solutions of the Hamiltonian
\begin{equation}\label{eq:b9}
H =
\begin{pmatrix}
E_{10} & t_c \\
t_c & E_{01}\\
\end{pmatrix}
\end{equation}
\\
\noindent with eigenvalues $E^\pm = 1/2(E_{10}+E_{01}) \mp 1/2 \sqrt{(E_{10}-E_{01})^2 + 4 t_c^2}$, as shown schematically in Fig.~\ref{Fig7}a. More generally, setting the electrochemical potential of the leads at $\mu=0$ and defining $\Delta E_1 = E_1 + (n+1/2)E_{C1} + mE_{Cm}$; and $\Delta E_2 = E_2 +(m+1/2)E_{C2} + nE_{Cm}$ this can be expressed as
\begin{widetext}
\begin{equation}\label{eq:b10}
E^\pm = E_{n,m}(E_1,E_2) + 1/2(\Delta E_1 + \Delta E_2) \mp 1/2 \sqrt{(\Delta E_1 - \Delta E_2)^2 + 4 t_c^2}
\end{equation}
\end{widetext}

\noindent The two wings observed in the stability diagram then correspond to $E^+ - E_{nm} = 0$ and $E_{n+1,m+1} - E^+ = 0$, respectively. By changing the variables to $\Delta = \Delta E_1 + \Delta E_2$ and $\epsilon = \Delta E_1 - \Delta E_2$  [see Fig.~\ref{Fig7}b], the separation between the wings $E_{\Delta}$ for a given detuning $\epsilon$ then follows from solving $E^+(\Delta,\epsilon) - E_{nm}(\Delta,\epsilon) = 0$ and $E_{n+1,m+1}(\Delta+E_{\Delta},\epsilon) - E^+(\Delta+E_{\Delta},\epsilon) = 0$, which yields

\begin{equation}\label{eq:b11}
E_{\Delta}=2(E_{Cm}+\sqrt{\epsilon^2+4t_c^2})
\end{equation}

\begin{figure}[b]
\includegraphics[width=70mm]{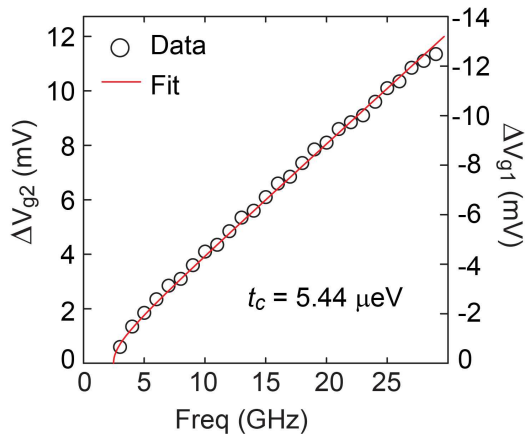}
\caption{\label{Fig9} Microwave-photon sideband spacing as a function of frequency. The solid red line is a fit to the data using Eq.~\eqref{eq:b12}.}
\end{figure}

\noindent Figure ~\ref{Fig8} shows the measured wing separation $E_{\Delta}$ as a function of detuning $\epsilon$ for three different charge configurations and top-gate voltages. The insets show the corresponding data in the stability diagrams as a function of $\epsilon$ and $\Delta$ using the conversion factors of Eq.~\eqref{eq:b8}. Fitting these curves to Eq.~\eqref{eq:b11}, we are able to extract $E_{Cm}$ and $t_c$ which are the fit parameters. Note that for the rightmost plot the tunnel coupling is very large and the system is effectively a single quantum dot. These measurements also demonstrate our ability to tune the coupling between the quantum dots.

For very small tunnel couplings, as for the data in the main text, it is difficult to accurately measure $t_c$ from the curvature of the wing in the stability diagram. Instead the tunnel coupling is determined more precisely from microwave data as in Fig.~\ref{Fig3}. This is further illustrated by Fig.~\ref{Fig9} which shows the position of the observed sidebands as a function of microwave frequency $f$ as the detuning is varied [e.g. along the dashed red line in the inset of Fig.~\ref{Fig2}b]. The vertical axes in Fig.~\ref{Fig9} show the gate-voltage spacings $\Delta V_{g1}$, $\Delta V_{g2}$ of the sidebands measured from the central resonance (i.e. zero detuning). We fit the data to

\begin{equation}\label{eq:b12}
\alpha \Delta V_{g2} = \sqrt{(hf)^2-4t_c^2}
\end{equation}

\noindent using $\alpha$ and $t_c$ as fit parameters, yielding $t_c=5.44$ $\mu$eV (or, expressed in frequency as in the main text, approximately 1.3 GHz). For the gate conversion factor we obtain  $\alpha \sim 0.01|e|$, which yields the $\epsilon = \alpha \Delta V_{g2}$ value used in Figs.~\ref{Fig2}c and ~\ref{Fig3}a - and which is consistent with the conversion factors measured from the stability diagrams [the precise values of $A-D$ in Eq.~\eqref{eq:b8} depend on the top- and side-gate voltages used to tune the dot-dot and dot-lead couplings; for the settings here, the best fits for $A-D$ yield  $0.0200; 0.0125; 0.0131; 0.0192$, respectively].\\
\\
The tunnel coupling can also be estimated from the \textit{width} of the (effective) $(0,1) \leftrightarrow (1,0)$ transition measured at zero detuning. The quantum capacitance is proportional to the curvature or second derivative of the energy dispersion in Fig.~\ref{Fig7}a for which we find

\begin{equation}\label{eq:b13}
\partial^2  E^+/\partial \epsilon^2 \propto \frac{4t_c^2}{(\epsilon^2 +4t_c^2)^{3/2}}
\end{equation}

\noindent which can also be expressed as a ratio of (Larmor) frequencies ${f_0}^2/f^3$, where we define $hf_0 = 2t_c$ and $hf = \sqrt{\epsilon^2 + 4t_c^2}$ as the energies of microwave photons to drive transition between the two quantum states at zero and finite detuning, respectively. Equation \eqref{eq:b13} yields a full width at half maximum (FWHM) of $\Delta E_{FWHM} = 4t_c \sqrt{2^{2/3}-1} \approx 3.07 t_c$. For a tunnel coupling $t_c=5.44$ $\mu$eV, we thus obtain $\Delta E_{FWHM} = 16.7$ $\mu$eV, which is consistent with the data [see e.g. Fig.~\ref{Fig3}a]. Since the width of the phase signal $\propto t_c$ and the magnitude of the phase signal $\propto 1/t_c$, it follows that the transition line is most visible when the tunnel coupling is small \cite{Chorley1, Cottet}.

\begin{figure}[b]
\includegraphics[width=0.33\textwidth]{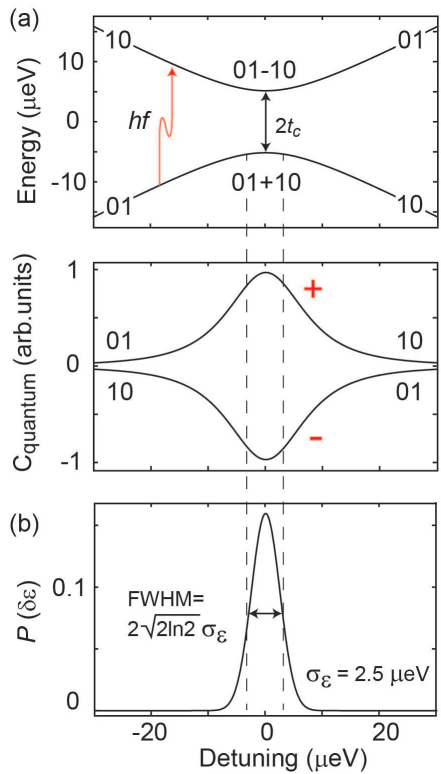}
\caption{\label{Fig10} (a) Energy band diagram (top panel) and quantum capacitance (bottom panel) as a function of level detuning. The indices denote the dot occupancies. (b) The noise in the detuning parameter is assumed to have a Gaussian distribution with a standard deviation $\sigma_{\epsilon}$.}
\end{figure}

\section{Dephasing and relaxation}
\label{appendix:c}

\subsection{Line shape at optimal point}
\label{appendix:c1}

\noindent The line shape shown in Fig.~\ref{Fig3}b is characterized by a pronounced asymmetry and a broad tail at the high-frequency side. This observation is consistent with dephasing at an optimal point where noise in the detuning parameter can only increase the energy splitting between the eigenstates. To understand and model the observed line shape, we make the following two assumptions:
\begin{itemize}
\item The charge relaxation time is longer than the rf measurement period $1/f_0$.
\item The charge noise is quasi-static and has a Gaussian distribution.
\end{itemize}

\noindent The first assumption implies that the ground- and excited-state populations do not relax into their equilibrium values during a rf measurement swing - on the order of 3 ns in our experiments. The observed phase shift due to the microwave drive will then be directly proportional to the quantum capacitances and time-averaged populations of the bonding and antibonding states \cite{Sillanpaa}. As illustrated in Fig.~\ref{Fig10}, the quantum capacitances of the bonding and antibonding states are detuning dependent and have opposite polarity.

\begin{figure*}[t]
\includegraphics[width=174mm]{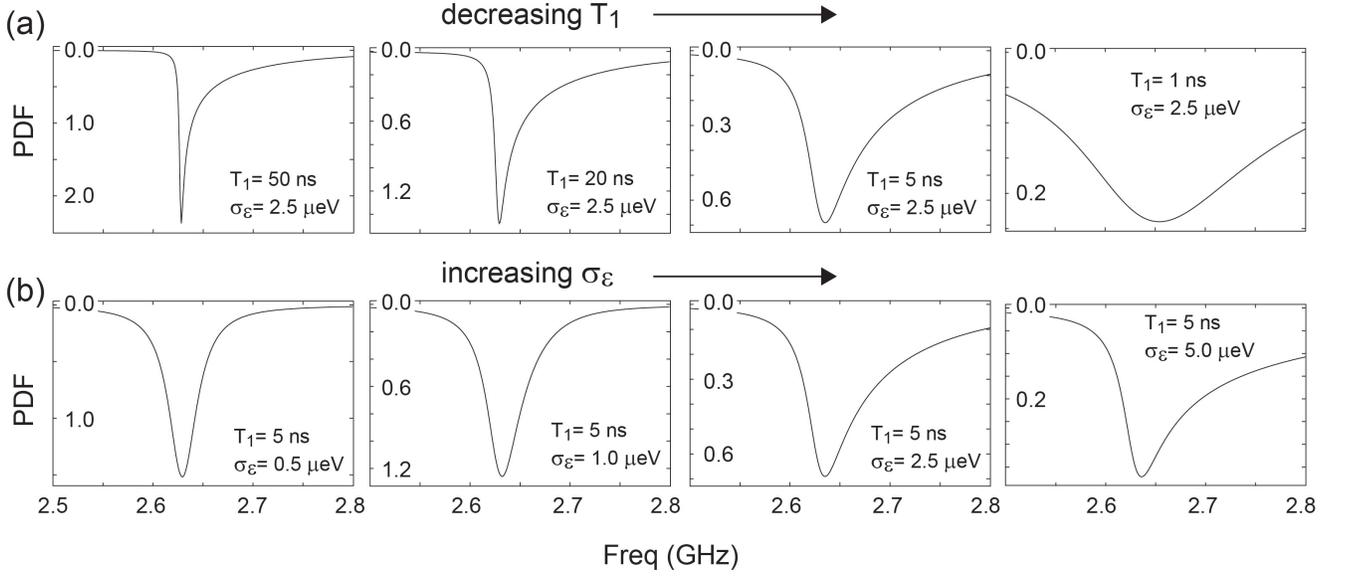}
\caption{\label{Fig11} Calculated probability density functions (PDFS) at an optimal point, as described in the text. The curves in (a) show calculations for decreasing relaxation times $T_1$, broadening the lines. The curves in (b) show calculations for different values of the standard deviation $\sigma_{\epsilon}$ of the noise distribution, which affects the high-frequency tails of the line shapes.}
\end{figure*}

\noindent The applied microwave signal drives transitions between the two quantum states, which are measured experimentally by the phase shift $\Delta \Phi$ of the reflected rf signal. The resonance line shape depends on the degree of charge relaxation and dephasing. In the Bloch equations, dephasing is taken into account by introducing a phenomenological parameter $T_2$ in the equations. Here, we take a different approach to include the dephasing that more accurately accounts for the low-frequency noise observed in the experiment \cite{Ithier}. To model the data analytically we assume the noise in the detuning parameter $\delta \epsilon$ to have a Gaussian distribution:
\begin{equation}\label{eq:c1}
P(\delta \epsilon) = \frac{1}{\sigma_{\epsilon}\sqrt{2 \pi}}e^{-\delta\epsilon^2/2 \sigma_{\epsilon}^2}
\end{equation}
\noindent where $\sigma_{\epsilon}$ is the standard deviation. Using the dependence of the energies of the states on detuning (see Fig.~\ref{Fig10}a) and assuming the noise to be quasistatic, we calculate the transformed probability density function as a function of energy, which is convolved with a Lorentzian of width $1/T_1$ - corresponding to an exponential decay factor $e^{-t/2T_1}$ in the time domain - to take charge relaxation into account. For continuous microwave excitation in the linear-response (or unsaturated) regime, the observed phase shift serves as a probe of this energy distribution. Figures ~\ref{Fig11}a and ~\ref{Fig11}b show how the line shapes vary with $T_1$ and $\sigma_{\epsilon}$, respectively. For all of the curves, we use a tunnel coupling $t_c = 5.44$ $\mu$eV, which corresponds to the experimental data, as described in Appendix \ref{appendix:b} above. As expected, increasing $\sigma_{\epsilon}$ has the effect of extending the high-frequency tail of the line shape while increasing relaxation broadens it.
\\

\begin{figure}[hb]
\includegraphics[width=0.30\textwidth]{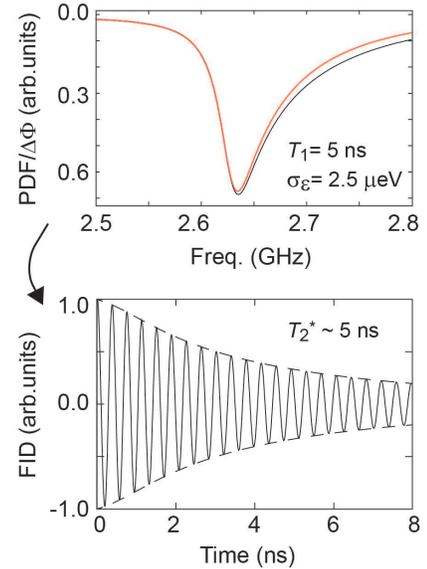}
\caption{\label{Fig12} (Top panel) Calculated probability density function (the black curve) and phase shift (the red curve) for $t_c = 5.44$ $\mu$eV, $\sigma_{\epsilon} = 2.5$ $\mu$eV and $T_1 = 5$ ns. (Bottom panel) Corresponding free-induction decay (FID), as in Eq.~\eqref{eq:c2}. $T_2^*$ is taken to be the time for which coherence has decayed to $1/e$ of its initial value.}
\end{figure}

\noindent To directly compare the calculated resonances with the experimental data - which measures phase shifts - we need to take into account that the quantum capacitance depends on detuning [see Fig.~\ref{Fig10}a] and decreases as $(f_0/f)^3$ for $f \geq f_0$ from its maximum value at zero detuning as described by Eq.~\eqref{eq:b13}. In addition, the magnitude of the observed phase shift will depend on the ratio of the driving amplitude and frequency \cite{Shevchenko}. As a result, the observed phase shift $\Delta \Phi \rightarrow 0$ for $f \gg f_0$, but is a small correction to the observed line shape in the frequency range of interest here; see Fig.~\ref{Fig12} (top panel, red line). We have also verified experimentally that the line shapes are not broadened by the rf probe signal. The amplitude of the rf signal at the device source electrode is difficult to determine precisely due to uncertainties in cable attenuation and reflections, but it is estimated to be less than -130 dBm. Given a resonator quality factor of about 30 and a lever arm of approximately 0.45, this amplitude yields a fluctuating detuning due to the rf signal $\delta \epsilon <1 \mu eV$.\\

\noindent The resonances shown in Fig.~\ref{Fig11} can also be understood as - that is, are equivalent to - the Fourier transform of a free-evolution decay (Ramsey-type) measurement where the time evolution is described by

\begin{equation}\label{eq:c2}
f(t) = \textrm{Re} \Biggl[\frac{e^{-(\frac{1}{2T_1}-i\frac{2t_c}{\hbar})t}}{\sqrt{1-\frac{i}{\hbar}\frac{\sigma_{\epsilon}^2}{2 t_c} t}}\Biggr]
\end{equation}
\noindent Here, the exponential decay factor $e^{-t/2T_1}$ is due to charge relaxation. The term in the denominator is due to pure dephasing and can be understood as follows: at the optimal point, the qubit is first-order insensitive to charge noise in the detuning parameter $\delta \epsilon$, and therefore second-order contributions dominate. The decay function due to charge noise is then described by the integral \cite{Ithier}

\begin{equation}\label{eq:c3}
f_{\textrm{decay}} (t) = \int d(\delta \epsilon) P(\delta \epsilon) e^{(\frac{i}{\hbar} \partial^2 E/2\partial \epsilon^2)\delta\epsilon^2 t}
\end{equation}

\noindent where $E=\sqrt{\epsilon^2+4t_c^2}$ is the energy difference between the qubit states. Assuming a Gaussian noise distribution $P(\delta \epsilon)$ [see Eq.~\eqref{eq:c1}] and given that at the optimal point $\partial^2 E/\partial \epsilon^2= 1/2 t_c$, this simplifies to

\begin{equation}\label{eq:c4}
\hspace{-25mm} f_{\textrm{decay}} (t) = \frac{1}{\sqrt{1-\frac{i}{\hbar}\frac{\sigma_{\epsilon}^2}{2 t_c} t}}
\end{equation}

\begin{figure*}
\includegraphics[width=160mm]{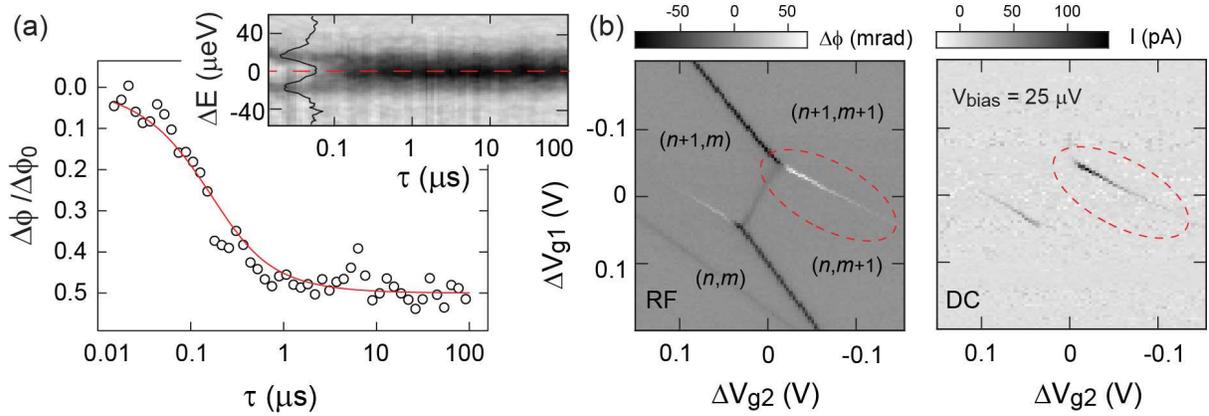}
\caption{\label{Fig13} (a) Normalized phase response measured at zero detuning in the presence of chopped microwaves with a 50\% duty cycle and a variable period $\tau$; see also Fig.~\ref{Fig3}c. (Inset) The phase response as a function of both $\tau$ and the detuning $\Delta E$. The line trace in the inset shows the response at $\tau = 20$ ns. (b) (Left panel) The rf phase response as a function of $V_{g1}$ and $V_{g2}$. (Right panel) The dc current as a function of $V_{g1}$ and $V_{g2}$ in the presence of a source-drain bias voltage $V_{bias} = 25$ $\mu$V for the same region as in (a). The dashed red lines in the panels indicate a region of cotunneling in the stability diagram.}
\end{figure*}

\noindent Dephasing is therefore dependent on the ratio $\sigma_{\epsilon}^2/{2 t_c}$, such that the charge qubit is less susceptible to charge noise for a larger $t_c$, but a larger $t_c$ also comes at the expense of a decrease in readout sensitivity as evident from Eq.~\eqref{eq:b13}. In our experiments, the standard deviation $\sigma_{\epsilon}$ can be estimated from the phase signal at finite detuning where there is an approximately linear $E-\epsilon$ relation. For a Gaussian distribution we expect the width to depend on the standard deviation, $2 \sqrt{2\ln2}\sigma_{\epsilon}$. The observed linewidths at finite detuning are on the order of $1.5 - 2.0$ GHz [see, e.g., Fig.~\ref{Fig3}a, right panel], which yields $\sigma_{\epsilon} \sim 2-3$ $\mu$eV. This standard deviation is consistent with the best fit that reproduces the line shape at the optimal point - that is, the data in Fig.~\ref{Fig3}b - which is obtained for $\sigma_{\epsilon} \sim 2.5$ $\mu$eV. Note that the amplitude of the phase signal is taken as an additional fit parameter. The deduced $T_2^*$ is taken to be the time for which coherence has decayed to $1/e$ of its initial value for these parameters, as shown in Fig.~\ref{Fig12}, bottom panel.\\
\\
\noindent The measured charge relaxation time $T_1 = 48 \pm 6$ ns obtained from the chopped microwave experiments as presented in Fig.~\ref{Fig3}c provides an upper bound for the $T_1$ parameter in the line-shape fits. This value is consistent with our findings. Figure ~\ref{Fig11}a shows how the calculated line shapes vary with a decreasing $T_1$, and a best fit to the experimental data yields a $T_1$ parameter in the (5-10)-ns range. We have verified that the measurements were not affected by the applied microwave power, which had been adjusted to values optimized for largest signal-to-noise ratio while low enough to not affect the intrinsic line shape in the linear-response regime. We speculate that the additional broadening (yielding a slightly lower $T_1$ in the line shape fits) is the result of a charge noise contribution to the tunnel-coupling parameter $t_c$ of order 0.1 $\mu$eV.

\subsection{Charge relaxation: rf and dc data}
\label{appendix:c2}

\noindent The charge relaxation time $T_1$ of our charge qubit is determined by using a chopped microwave signal with 50 \% duty cycle and variable period $\tau$. For short periods where $\tau \ll T_1$, the system has no time to relax during the second half of a period when the microwave signal is switched off. In this case, the phase signal at zero detuning is close to the background value, as apparent from the data in the inset of Fig.~\ref{Fig13}a which shows the phase response as a function of $\tau$ and the detuning $\Delta E$. As $\tau$ is increased, the signal shows an exponential decay, approaching $\Delta \Phi(\tau)/\Delta \Phi_0 \rightarrow 1/2$ when $\tau \gg T_1$, as shown in Fig.~\ref{Fig13}a - reproduced from Fig.~\ref{Fig3}c.\\
\\
\noindent We believe that the limiting factor for $T_1$ in our device is tunneling to the leads. Figure \ref{Fig13}b shows both the phase response (left panel) and dc transport data (right panel) in the stability diagram. Cotunneling is evident in the dc transport data as an extended current region along the horizontal resonances, as indicated by the dashed red lines, with the magnitude decreasing as the distance from the triple points increases. In the rf data, cotunneling is evident as a phase shift to positive values (the 'white' lines in the rf data). While we do not attempt to provide a quantitative analysis of cotunneling here, we note that a relaxation time of $T_1 = 48 \pm 6$ ns would correspond to rates of about 20 MHz. This value is indeed consistent with the magnitude of the dc current in the (1-5)-pA range observed in the data (away from the triple points). Relaxation via the leads (electron exchange) is thus the most likely limiting factor for $T_1$ in our device.



\end{document}